\documentclass[preprint]{aastex}
\usepackage{epstopdf}
\def\apgt{\ {\raise-.5ex\hbox{$\buildrel>\over\sim$}}\ }
\def\aplt{\ {\raise-.5ex\hbox{$\buildrel<\over\sim$}}\ }
\def\Msun{M_\odot}

\shorttitle{Non-equipartition in Novae}
\shortauthors{Shara et al.}

\begin{document}

\title{Non-Equipartition of Energy, Masses of Nova Ejecta, and Type Ia Supernovae}

\author{
Michael~M.~Shara\altaffilmark{1}
Ofer~Yaron,\altaffilmark{2}
Dina~Prialnik,\altaffilmark{2}
Attay~Kovetz,\altaffilmark{2,3}
}

\altaffiltext{1}{Department of Astrophysics, American Museum of Natural
History, Central Park West and 79th street, New York, NY 10024-5192}
\altaffiltext{2}{Department of Geophysics and Planetary Sciences,
Sackler Faculty of Exact Sciences, Tel Aviv University, Ramat Aviv
69978, Israel}
\altaffiltext{3}{School of Physics and Astronomy, Sackler Faculty of
Exact Sciences, Tel Aviv University, Ramat Aviv 69978, Israel}

\begin{abstract}

The total masses ejected during classical nova eruptions are needed to answer two questions with broad astrophysical implications: Can accreting white dwarfs be pushed towards the Chandrasekhar mass limit to yield type Ia supernovae? Are Ultra-luminous red variables a new kind of astrophysical phenomenon, or merely extreme classical novae? We review the methods used to determine nova ejecta masses. Except for the unique case of BT Mon (nova 1939), all nova ejecta mass determinations depend on untested assumptions and multi-parameter modeling. The remarkably simple assumption of equipartition between kinetic and radiated energy ($E_{\rm kin}$ and $E_{\rm rad}$, respectively) in nova ejecta has been invoked as a way around this conundrum for the ultra-luminous red variable in M31. The deduced mass is far larger than that produced by any classical nova model. Our nova eruption simulations show that radiation and kinetic energy in nova ejecta are {\it very far} from being in energy equipartition, with variations of four orders of magnitude in the ratio $E_{\rm kin}/E_{\rm rad}$ being commonplace. The assumption of equipartition must not be used to deduce nova ejecta masses; any such ``determinations" can be overestimates by a factor of up to 10,000.
We data-mined our extensive series of nova simulations to search for correlations that could yield nova ejecta masses. Remarkably, the mass ejected during a nova eruption is dependent only on (and is directly proportional to) $E_{\rm rad}$. If we measure the distance to an erupting nova and its bolometric light curve then $E_{\rm rad}$ and hence the mass ejected can be directly measured. 
\end{abstract}


%% --- section ---

\section{Motivation}
\label{introduction}

\subsection{SNIa}

Which type of star (or stars) is the progenitor of type Ia supernovae (SNIa)? The discovery, via SNIa, of the likely acceleration of the expansion of the universe inexorably drives us to the concept of dark energy. If correct, this breakthrough in our understanding of cosmology is profound... but significant uncertainties remain. One of the most perplexing of these uncertainties is that we still can't identify the progenitors of type Ia supernovae. Systematic changes in these progenitors over a Hubble time could systematically affect SNIa light curves in ways we can't predict; a worrying prospect for the brightest standard candles known. There is a broad consensus that if a white dwarf (WD) can somehow be made to accrete enough matter to exceed the Chandrasekhar mass, it will fuse carbon and explode, releasing $ \sim 10^{51} $ ergs in photons and radioactive nickel-56 (observationally-constrained quantities). The problem, of course, is that there are multiple possible donors able to dump matter onto a white dwarf in a binary system: brown dwarfs, white dwarfs, main sequence stars, red giants and supergiants. The absence of hydrogen in SNIa spectra is an important constraint, but unfortunately it doesn't definitively rule out any of the donor suspects. Just a tiny amount of accreted matter ($\sim 3\times10^{-10}~M_\odot$~yr$^{-1}$)) - with or without hydrogen - can push a nearly-Chandrasekhar mass WD ``over the edge". Thus none of the hydrogen-rich donors noted above are excluded, at least in principle, if the accreting WD gains mass in secular fashion.

WDs accreting hydrogen-rich matter usually undergo classical nova (CN) eruptions, which periodically eject mass. Knowing whether such WDs increase or decrease in mass as a result of these competing processes is equivalent to knowing if WDs in mass-transferring binary systems, with hydrogen-rich secondaries, can be the progenitors of SNIa.  An effective way to determine if the mass of an accreting WD is secularly increasing would be to measure its rate of accretion and time between eruptions. The accreted mass must then be compared to the mass ejected ($m_{\rm ej}$) to determine if there is a net gain or loss in mass. Recent attempts to carry out this experiment with the recurrent nova (RN) T Pyx, and the difficulties with directly measuring $m_{\rm ej}$,  are reported in \citet{sch09} and \citet{sel08}.

T Pyx is the best studied and prototypical RN. RN might give rise to SNIa; the importance of knowing whether T Pyx will (or will not) become an SNIa cannot be overstated. \citet{sch09} have mustered strong evidence that T Pyx must have undergone a {\it classical} nova eruption around 1866, after a 2.6 Myr hibernation episode \citep{sha86} followed by a 750 Kyr phase of accretion at a rate of $\sim 4\times10^{-11}~M_\odot$~yr$^{-1}$. Allowing for surrounding interstellar matter swept up by the expanding nova envelope, the 1866 classical nova ejecta mass should now total at least $750\  {\rm Kyr}\times\sim 4\times10^{-11}~M_\odot$~yr$^{-1}$ $\sim 3\times 10^{-5} \Msun$. The mass accreted and ejected since 1866, in five recorded recurrent nova outbursts, must be a few orders of magnitude less. If this scenario is correct then T Pyx is unlikely to become a SNIa, because the WD mass is secularly decreasing due to classical nova eruptions. A direct test of this scenario - and whether {\it the} prototypical recurrent nova will become an SNIa - would be an unambiguous measurement of the mass in the ejecta surrounding T Pyx. A direct measurement of the total mass surrounding T Pyx is very challenging, as successive generations of ejecta catch up with and shock gas already in place. Some shocked blobs appear and disappear on a timescale as short as a year \citep{sha86}.

\subsection{Ultra-luminous Red Novae}

Over the past decade it has been suggested that a few rare, but extraordinarily luminous eruptive variables are prototypes of a new class of astrophysical phenomenon \citep{har81}. The ``Red Variable" in M31 \citep{rmp89} and \citep{mou90}, V838 Mon \citep{bon03} and the ``Optical Transient" in M85 \citep{kul07} and \citep{rau07} reached absolute magnitudes as luminous as $-10$ to $-12$. This is brighter than classical novae, but fainter than supernovae.

Classical novae typically eject $\sim 1-10\times 10^{-5} \Msun$ during an outburst \citep{yar05}. It is evident that mass ejections at least 100 times larger must have accompanied the outbursts of the Ultra-luminous Red Novae, as the ejecta remained optically thick to much larger distances from the sites of the explosions. As a result, the ejecta cooled to temperatures as low as 700 Kelvins. Explaining these observations is challenging. Some of the astronomical community supports the hypothesis of ``mergebursts" (the merger of a binary star) \citep{sok03} and \citep{tyl06}. The mass ejected in such an event could easily exceed $\sim 10^{-2} \Msun.$ A competing model \citep{sha10} posits that extreme classical novae (with very low ($0.5 \Msun$)) WD masses and very high accreted envelopes ($\sim 10^{-3} \Msun$) can explain the Ultra-luminous red variables without invoking a new type of astrophysical phenomenon.  An important numerical prediction of the extreme nova scenario is that the mass ejected in such an event is at most  $\sim 2-3\times 10^{-3} \Msun$. Thus a direct measurement of $m_{\rm ej}$ in excess of $\sim 10^{-2} \Msun$ could eliminate one of the competing explanations for Ultra-luminous red variables.

%% --- section ---
\section{Measuring Ejecta Masses}

The preceding section emphasizes how desirable it would be to accurately measure the masses of ejecta from classical novae. The ejected masses of classical novae have long been debated and estimates have widely varied, depending on the methods employed to interpret observations. Estimates of the density of ejected gas (from the presence or absence of forbidden emission lines) and size of the ejecta (from the angular size and velocity of the ejecta) lead to the simplest possible estimated ejecta masses. However, the oft-made simple assumptions of homogeneous, spherically symmetric mass ejection are demonstrably false \citep{sha97}, \citep{wad00}. Filling factors very different from unity are probably common, as are dense clumps in outflowing nova winds \citep{wil04}. \citet{fer98} has emphasized that nova ejecta masses are systematically underestimated because gas emitting in any particular region of the electromagnetic spectrum (say the ultraviolet) can have very different properties from gas emitting in another region (say the thermal infrared). A region of gas may be an efficient emitter in one regime and totally invisible in another. These factors can easily lead to uncertainties of orders of magnitude in the estimated ejected masses.

Could any or all of these factors - clumpiness, variable filling factors, variable ionization, and especially non-spherical ejection, - render all of the detailed 1-D nova models too doubtful to trust? While highly non-spherical ejection might compromise the 1D models, the other factors are challenges in interpreting the observations rather than the models. Fortunately there is one very dependable measurement of a nova's ejected mass, which serves as a critical ``sanity check" for extensive simulations of nova eruptions \citep{yar05}. \citet{sch83} measured the period change of the eclipsing classical nova BT Mon (nova 1939) as a result of mass loss from the binary system. The modern period of this binary is 40 parts per million longer than before 1939. Reasonable assumptions concerning the angular momentum carried off by the ejected material provide a dynamical measure of the ejecta mass. The mass of the WD in BT Mon has been measured with rather high precision: $1.04 \pm 0.06 \Msun$ \citep{smi98}.

Repeating \citet{sch83}'s derivation but with the more precise white dwarf mass determinations of \citet{smi98} yields the BT Mon ejected mass: $\sim 4\times 10^{-5} \Msun$, to within a factor of two or so. A heroic reconstruction of the light curve of BT Mon from Harvard archival photographic plates \citep{sch83} suggests a time to decline, by two ($t_2$) and three ($t_3$) magnitudes, of 140 and 190 days, respectively.
If correct, these would be useful constraints for testing nova simulations. Unfortunately, the maximum light of BT Mon was almost certainly missed (when the star was in the daytime sky), based on the outburst spectra \citep{whi40}, \citep{san40}, \citep{swi41}. These spectra demonstrate that the outburst must have occurred when the nova was at least $5.4$ magnitudes brighter than when the photographic record began. Thus the $t_2$ and $t_3$ above do not apply to maximum light; the correct values must be considerably smaller, but we can't measure them directly. Fortunately though, the full widths of the emission lines {\it were} well observed, corresponding to velocities of 1500 km/s \citep{whi40}, 2100 km/s \citep{san40} and 1730 km/s \citep{swi41}. We adopt the maximum ejected velocity as half the average of these three observations, viz. 900 km/s.

Each of the nova simulations of \citet{yar05} predicts an ejected mass and a maximum ejection velocity for a given white dwarf mass, mass accretion rate and luminosity. These quantities vary quite smoothly between simulations, so it is easy to interpolate. It is reassuring that, for a cold WD (10 Million Kelvins core temperature) of $1.0 \Msun$, accreting at a rate of $\sim 3\times10^{-10}~M_\odot$~yr$^{-1}$, the 1-D nova simulation grid of \citet{yar05} predicts an ejected mass of $\sim 7\times 10^{-5} \Msun$ and a maximum ejection velocity $v$ of 1000 km/s, in very good agreement with the observations. (A further check would be possible if the WD luminosity were directly measureable, but because of ongoing accretion this isn't possible. There is also no easy way to check on the BT Mon WD core temperature, but 10 Million Kelvins is certainly plausible). Unfortunately not a single other dynamical mass for nova ejecta has ever been measured, so the ejected masses of Ultra-luminous red variables, recurrent novae and of all other novae remain uncertain.

This uncertainty is overcome if one assumes equipartition of energy between radiation and motion (e.g., \citet{mou90}) in nova ejecta. As the total radiated energy $E_{\rm rad}$ is readily obtained by integrating the light curve, and an average expansion velocity $v_{\rm av}$ may also be obtained from spectrographic observations, the ejected mass is derived: $m_{\rm ej}\sim2E_{\rm rad}/v_{\rm av}^{2}$. If the assumption of equipartition of energy could be trusted then a simple and powerful tool would be available for testing nova theory, and ideas about SNIa progenitors.

\section{Non-Equipartition Between Kinetic and Radiated Energy }

We decided to test this {\it ad-hoc} and very important assumption via our extensive evolutionary calculations of classical novae. The hydrodynamic, 1-D Lagrangian stellar evolution code used in all our studies is described in some detail \citep{pkz95}. It includes OPAL opacities, an extended nuclear reactions network comprised of 40 heavy element isotopes, and a mass-loss algorithm that applies a steady, optically thick supersonic wind solution (following the phase of rapid expansion). In addition, diffusion is computed for all elements, accretional heating is taken into account and convective fluxes are calculated according to the mixing length theory. The code finely subdivides the outer layers so that the radius at which optical depth becomes unity is well determined, as is the effective temperature. This, in turn, allows the calculation of the radiated energy at each time-step.

Initial models were prepared for a range of WD-mass values and three temperatures by cooling WD models from higher temperatures. Each nova model was followed through several consecutive outburst cycles in order to eliminate the effect of the initial configuration. One typical cycle was then chosen as representative of each parameter combination. The kinetic energy $E_{\rm kin}$ in each simulation was obtained by summing ${1\over2}\dot M_{\rm ej}\delta t v^{2}$, where $\dot M_{\rm ej}$ is the mass ejection rate, over the entire mass loss episode. Similarly the total radiated energy $E_{\rm rad}$ was computed by summing the radiation emitted at all time-steps from the  model's photosphere, $L\delta t$.

\begin{figure}
\plotone{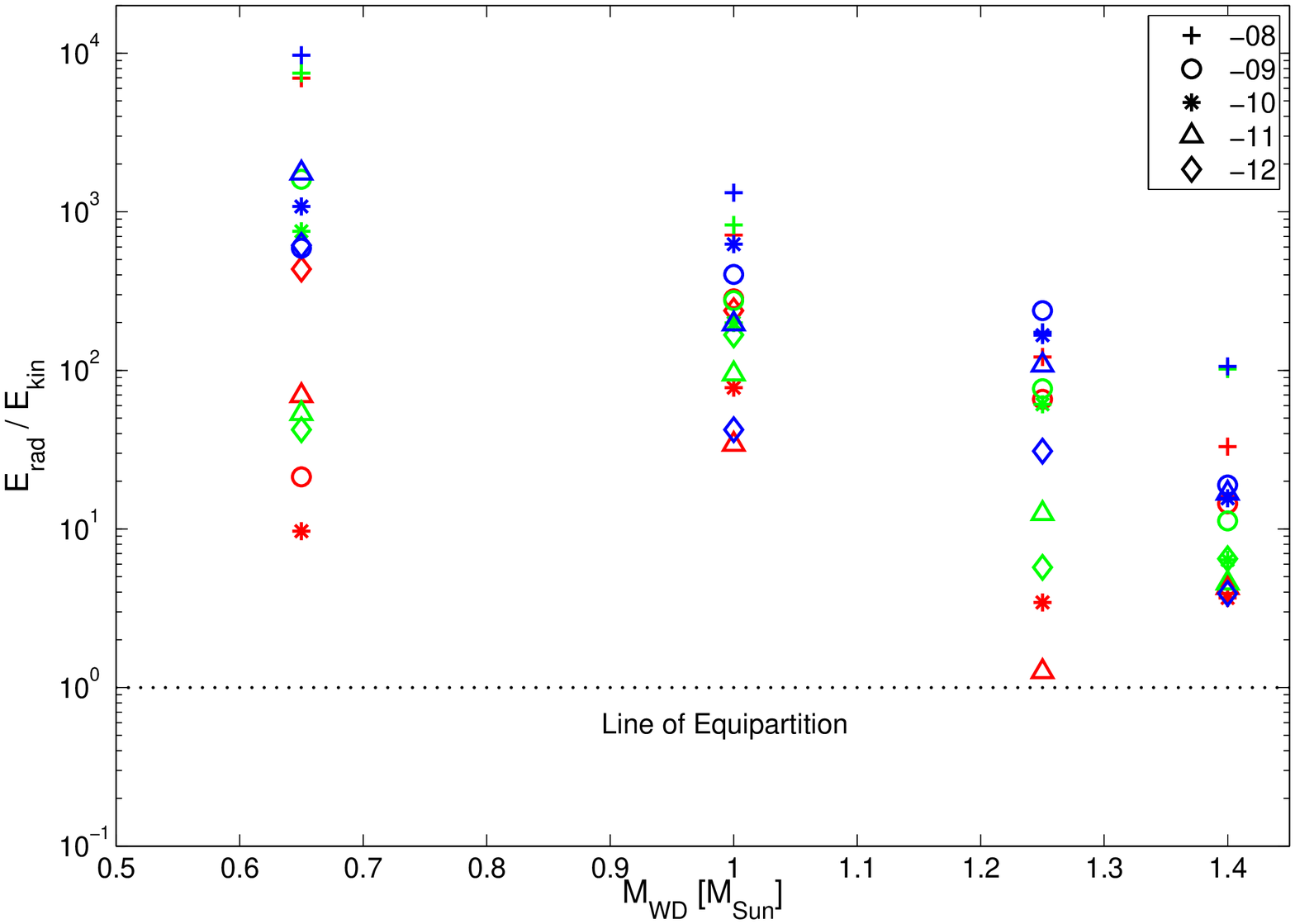}
\caption{Ratio of radiated to kinetic energy of a nova outburst obtained
from 58 model calculations for several parameter combinations, as marked. Different symbols correspond to different accretion rates, given in the legend in units of 
$\log \dot M$ $(M_\odot\ {\rm yr}^{-1})$. 
Red, green and blue symbols correspond to core WD temperatures $T_{\rm WD}$ of 10, 30 and 50~$\times10^6$ Kelvins.}  
\label{fig:eradekin}
\end{figure}

Results for 58 simulations, using the full range of nova white dwarf masses, temperatures and accretion, including the models of \citet{yar05} and \citet{sha10} are shown in Figure~\ref{fig:eradekin}; $E_{\rm rad}/E_{\rm kin}$ values of 10 to 1000 are the norm. But we see that the ratio of radiated to kinetic energy can be as small as 1 or large as 10,000!  Confirming this, a single, very detailed simulation of a solar mass white dwarf yielded a ratio of 10 \citep{kov99}. Figure~\ref{fig:eradekin} demonstrates that $E_{\rm rad}/E_{\rm kin}$ tends to decrease with increasing WD mass, but we need more information -- the underlying WD luminosity {\it and} accretion rate -- to uniquely determine the value of $E_{\rm rad}/E_{\rm kin}$.  If we could somehow measure the WD mass, its intrinsic luminosity, and the accretion rate before the eruption we could then use Figure~\ref{fig:eradekin} to uniquely determine $E_{\rm rad}/E_{\rm kin}$. This information is simply not available. If it were, we could in fact determine the ejected mass directly from the grid of models.

Mass estimates that simply assume that $E_{\rm rad}/E_{\rm kin}\ =\ 1$ will be in error by up to a factor of 10,000. {\it Thus, assuming equipartition results in gross overestimates of the ejected masses derived from observations of ejection velocities and nova luminosities}. We cannot emphasize too strongly that the assumption of equipartition between luminosity and motion in nova ejecta is unjustified and leads to extremely {\it overestimated} masses. This may help settle the controversy between the low ejecta masses advocated by nova modelers and the (sometimes) significantly higher masses claimed by observers. In particular, equipartition-based suggestions of $10^{-2}\Msun$ or even $10^{-1}\Msun$ for the ejecta of objects like M31-RV should be discounted.

Even though the assumption of equipartition fails in predicting $m_{\rm ej}$, we are not forced to end the discussion on a negative note. In the following section we demonstrate that nova ejecta masses can be determined with readily available observational data.

\section{Nova Ejecta Masses from Ejection Velocities and $E_{\rm rad}/E_{\rm kin}$}
%\subsection{Which Nova parameters control $E_{\rm rad}/E_{\rm kin}$ ?}

In principle, if $E_{\rm rad}\ne E_{\rm kin}$, the nova ejecta mass may still be obtained from
\begin{equation}
m_{\rm ej} = [ E_{\rm rad}/ (0.5 v^2)] [E_{\rm kin}/E_{\rm rad}] ,
\label{eq:mej}
\end{equation}
where the first term is obtained purely from observations and the second term is provided by models. 

The (distance-independent) quantity $v$ requires a series of spectra of a nova to determine $v$. 
Measuring $E_{\rm rad}$ itself can be more challenging: an accurate distance to a nova and its bolometric light curve are required. The distance can be determined from the absolute magnitude - decline time relationships of \citet{sha81} and \citet{dow00}, or the expansion parallax of the nova shell, or the distance of a host star cluster or galaxy in which a nova is located. In any such derivation it is essential to include the very large contributions of ultraviolet and infrared radiation from novae, typically seen after the peak in visual brightness \citep{gal74}, \citep{gal76}, \citep{gei70}, \citep{geh80}. 

\begin{figure}
\plotone{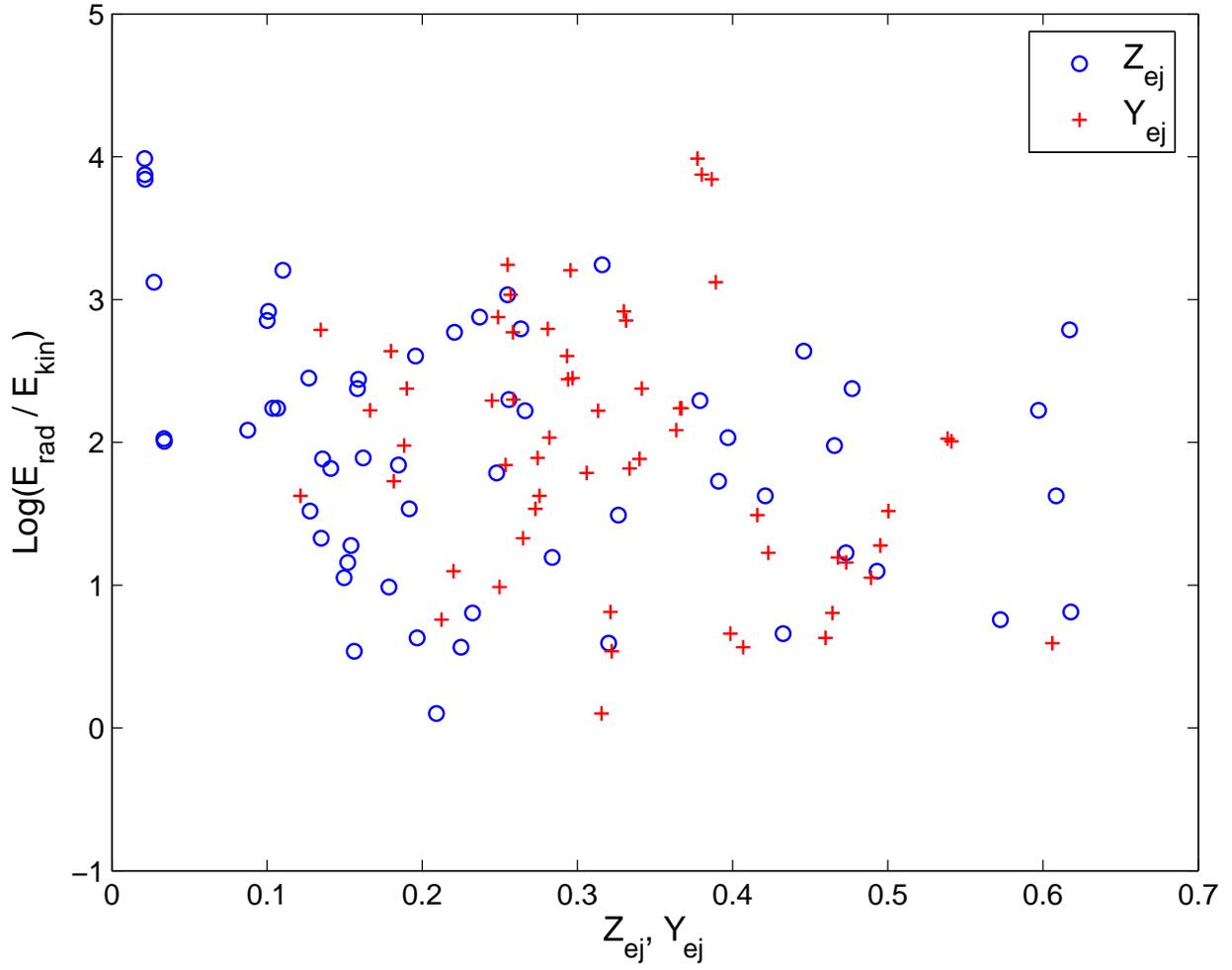}
\caption{The ratio of kinetic to radiated energy in 58 nova models as functions of helium abundance $Y$ and metallicity $Z$. No correlation is seen. }
\label{fig:YZ}
\end{figure}

The problem is how to connect models with observations for as few observables as possible.
As shown by \cite{pkz95}, nova outbursts constitute a 3-parameter family of events. Three independent observable parameters that span the parameter space may be taken to be the helium abundance $Y$ and the metallicity $Z$ in a nova's ejecta, and the time of decline of the nova luminosity $t_3$ (or some other related observable like ejection velocity \citep{sha81}).   
These parameters, the composition-related ones in particular, are not always easy to determine.
Do $Y$ or $Z$ in a nova's ejecta control the value of $E_{\rm rad}/E_{\rm kin}$? In Figure~\ref{fig:YZ} we have plotted $E_{\rm rad}/E_{\rm kin}$ versus $Y$ and $Z$ for 58 nova models. The resulting scatter-plot tells us clearly that neither $Y$ nor $Z$ correlate with $E_{\rm rad}/E_{\rm kin}$. The correlation with $t_3$ is slightly better but not at all convincing.

Is there some other observable parameter that correlates with $E_{\rm rad}/E_{\rm kin}$?
As radiation pressure is the main driver of mass loss during a nova eruption, we might expect the mass loss rate $\dot M_{\rm ej}$ to behave similarly to Reimers's mass loss formula that is used in conjunction with red giant winds: $\dot M_{\rm ej}\propto LM/R$. Since the photospheric radius and WD mass change little during the mass loss episode of a nova, it follows that $\dot M_{\rm ej}$ might be roughly proportional to the luminosity. If this were the case, then the total ejected mass $m_{\rm ej}$ should be proportional to the product of a nova's luminosity and duration of mass ejection $ \sim  E_{\rm rad}$. Combined with  $E_{\rm kin} = 0.5 m_{\rm ej}v^2$, these relationships suggest that  $E_{\rm rad}/E_{\rm kin}$ should be proportional to $v^{-2}$.
%\subsection{Erad/Ekin and v}

\begin{figure}
\plotone{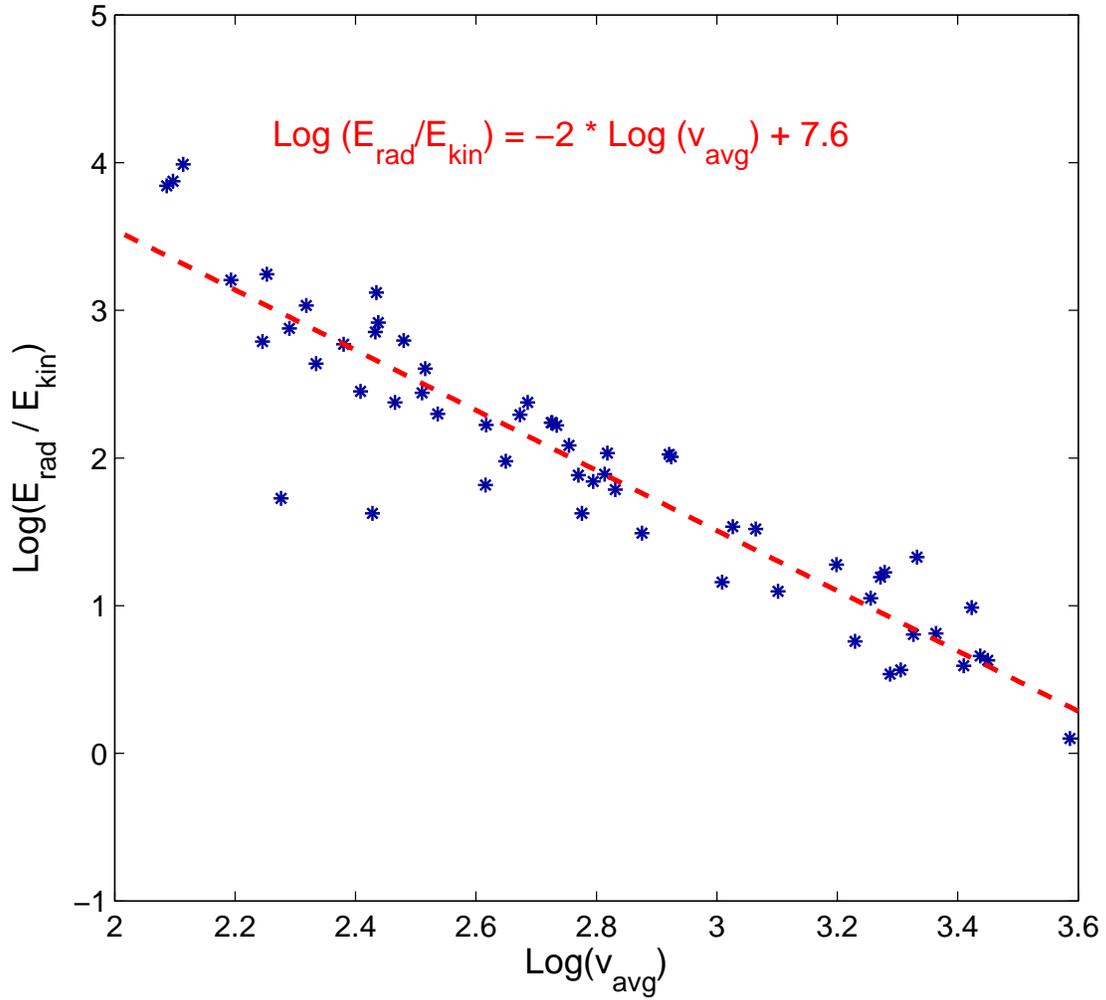}
\caption{The ratio of kinetic to radiated energy in 58 nova models as a function of average ejected velocity. The clear correlation is discussed in the text, and is key to determining the masses of nova ejecta. }
\label{fig:eradekinv}
\end{figure}

Our nova simulations can be used to test this heuristic suggestion. Figure~\ref{fig:eradekinv}, using the same 58 simulations of Figure~\ref{fig:eradekin}, shows the above suggestion to be correct, with remarkably small scatter, when we plot $E_{\rm rad}/E_{\rm kin}$ versus the average ejection velocity $v$: 
\begin{equation}
\log(E_{\rm rad}/E_{\rm kin}) = -2.0 \log (v) +7.6  
\label{eq:ratio}
\end{equation}
where the average ejection velocity $v$ is measured in km/s .
Equations (1) and (2)  immediately lead to
\begin{equation}
m_{\rm ej} =  6\times10^{-18} E_{\rm rad}
\label{eq:mejnet}
\end{equation}
where the ejected mass is measured in gm and the radiated energy in ergs. Figure 4 is a plot of the 58 models' 
computed ejected mass versus radiated energy. The line through the data points in Figure 4 is equation 3, the key result of this paper.

We thus derive the remarkable result that the mass of a nova's ejecta can be determined to within a factor of two (as seen from Figure 4) if we can measure just the time-integrated bolometric luminosity; spectra to determine $v$ are not needed. It is still a significant task to determine both the distance and especially the UV through IR flux of a nova, but this information alone is sufficient to determine the long-sought masses of nova ejecta. The coefficient in equation 3 depends on our grid of nova models; its precise value should eventually be directly measured in a few novae with well-determined ejecta masses and total radiated energies.

Unfortunately, BT Mon and its dynamically determined shell mass cannot be used to test this methodology. The nova reached peak brightness when it was in the daytime sky, it had faded at least five magnitudes before it was discovered \citep{whi40}, which was decades before IR or UV measurements were possible. Thus the total radiated energy (including the contributions to $E_{\rm rad}$ from the emitted UV and IR radiation) are unknown. Still, we point out that there is reason to be optimistic for the future. \citet{sch83} succeeded in measuring the BT Mon period change because of the treasure trove of plates in the Harvard plate stacks. It is likely that novae will be imaged many times, before they erupt, during the course of coming synoptic surveys like Pan-Starrs and LSST. We can therefore expect that we will eventually acquire period changes and dynamical masses for more classical and recurrent nova ejections. Such data will be invaluable to test the methodology proposed in this paper for measuring the ejected masses of all novae with well measured distances and UV through IR light curves, and especially to help constrain models of SNIa progenitors.

\section{Conclusions}

We have demonstrated that the assumption of equipartition of energy between radiation and kinetic energy in nova ejecta is incorrect.
The ratio $E_{\rm rad}/E_{\rm kin}$ for novae varies by four orders of magnitude. Any nova masses derived under the assumption of equipartition of energy may thus be wrong by up to a factor of 10,000. We find that there is no correlation between $E_{\rm rad}/E_{\rm kin}$ and either $Y$ or $Z$. However, $E_{\rm rad}/E_{\rm kin}$ is well-correlated with the average velocity of ejection $v$ during a nova eruption. This leads to the remarkable result that the total ejected mass from a nova is directly proportional to, and dependent only on, the total bolometric energy radiated by that nova. A simple formula for the ejected mass in a nova explosion is now available if the nova distance and bolometric light curve can be measured.

\acknowledgements
We thank an anonymous referee for excellent suggestions.

\pagebreak

\begin{figure}
\plotone{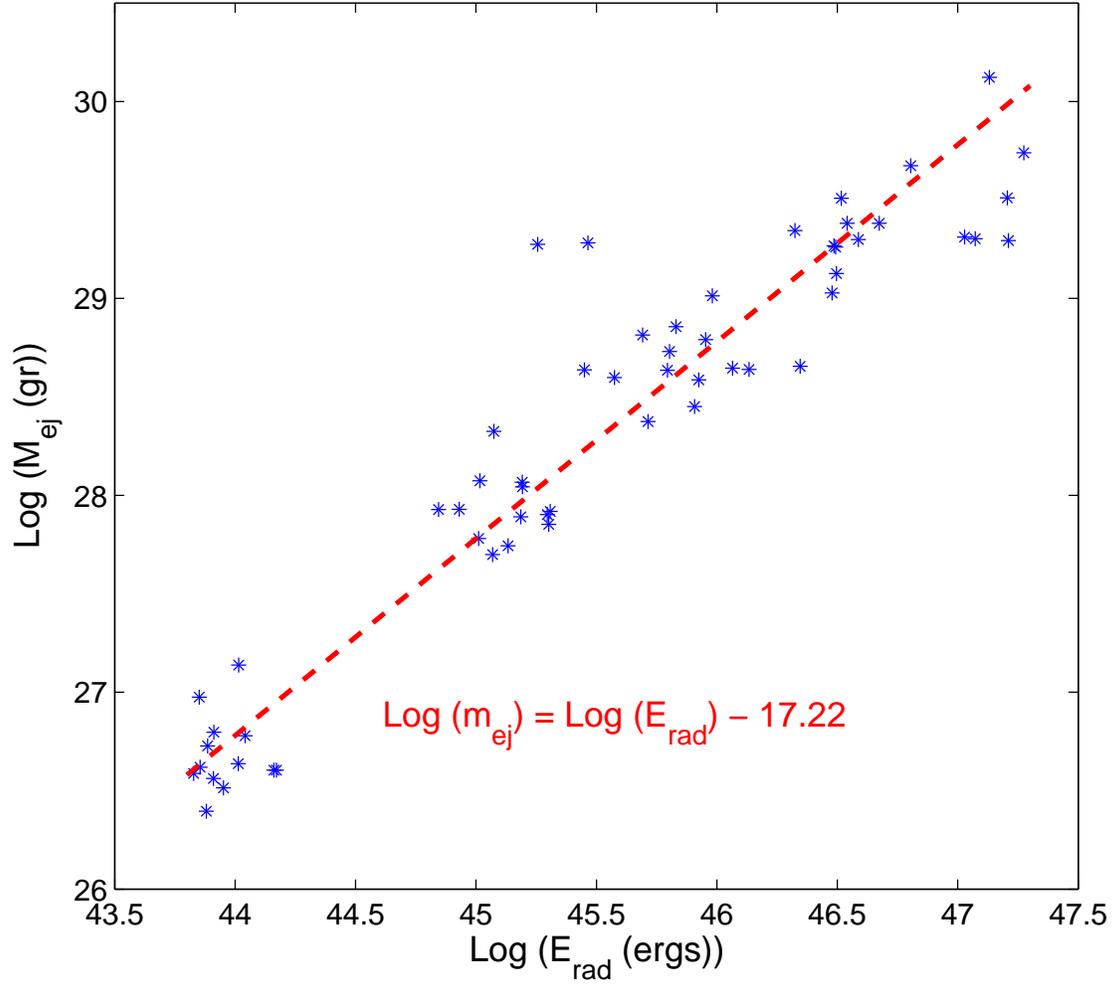}
\caption{The ejected mass as a function of radiated energy in 58 nova models. The clear correlation is discussed in the text; this is our key prediction for determining the masses of nova ejecta. }
\label{fig:mejerad}
\end{figure}

\end{document}